\documentclass[prl,aps,twocolumn,showpacs,superscriptaddress,floatfix]{revtex4}
\usepackage{graphicx}
\usepackage{bm}

\usepackage[dvips]{color}

\begin{document}

\title{Exact results for intrinsic electronic transport in graphene}

\author{Shijie Hu}
\affiliation{Department of Physics, Renmin University of China,
Beijing 100872, China} \affiliation{Institute of Theoretical
Physics, CAS, Beijing 100080, China}
\author{Wei Du}
\affiliation{Department of Physics, Renmin University of China,
Beijing 100872, China}
\author{Guiping Zhang}
\affiliation{Department of Physics, Renmin University of China,
Beijing 100872, China}
\author{Miao Gao}
\affiliation{Department of Physics, Renmin University of China,
Beijing 100872, China}
\author{Zhong-Yi Lu}
\affiliation{Department of Physics, Renmin University of China,
Beijing 100872, China}
\author{Xiaoqun Wang}
\affiliation{Department of Physics, Renmin University of China,
Beijing 100872, China}

\date{\today}
\begin{abstract}

We present exact results for the electronic transport properties of
graphene sheets connected to two metallic electrodes. Our results,
obtained by transfer-matrix methods, are valid for all sheet widths
and lengths. In the limit of large width-to-length ratio relevant to
recent experiments, we find a Dirac-point conductivity of $2e^2/\sqrt{3}h$
and a sub-Poissonian Fano factor of $2 - 3\sqrt{3}/\pi \simeq 0.346$ for
armchair graphene; for the zigzag geometry these are respectively 0 and 1.
Our results reflect essential effects from both the topology of graphene
and the electronic structure of the leads, giving a complete microscopic
understanding of the unique intrinsic transport in graphene.

\end{abstract}


\maketitle

Graphene, a graphite monolayer of carbon atoms forming a honeycomb
lattice, has a distinctive electronic structure whose low energy
excitations are described by massless Dirac fermions. The successful
extraction of micron-scale graphene sheets from a natural graphite
crystal, and their deposition onto an oxidized Si wafer \cite{exp0},
was a truly seminal event which ushered in a new era of realistic
experimental and theoretical exploration. The subsequent explosion
of graphene activity has focused on fundamental questions concerning
the transport properties of relativistic particles in graphene and on
its potential applications as a high-mobility semiconductor.

Theoretical predictions \cite{fermion} for two-dimensional Dirac-fermion
systems give an intrinsic conductivity $\sigma_0$ of order $e^2/h$. Minimal
conductivities around this value were observed at the Dirac point \cite{exp1}
in Ref.~\cite{exp0}, while later measurements \cite{exp3} suggested that
$\sigma_0 \rightarrow 4e^2/\pi h$ when the width-to-length ratio of the
sample is sufficiently large. This is the value obtained using massless
Dirac fermions and graphite leads in a Landauer-B\"{u}ttiker (LB) formulation
\cite{Landauer1,Landauer2}. It is associated with a maximum of 1/3 in the
Fano factor, $F_0$ \cite{Landauer2}, which reflects the partial transmission
of quantized charge through the finite graphene system. Measurements of the
current shot noise in both ballistic \cite{fano1} and diffusive \cite{fano2}
graphene systems have indeed found that $F_0 \approx 1/3$ in short and wide
samples. Many authors have addressed different aspects of the graphene
transport problem, which we summarize below. While the finite conductivity
and suppressed Fano factor are generally expected in graphene systems, the
underlying physics remains rather poorly understood, not least because the
carrier density at the Dirac point is zero.

\begin{figure}[t]
\includegraphics[width=7.7cm]{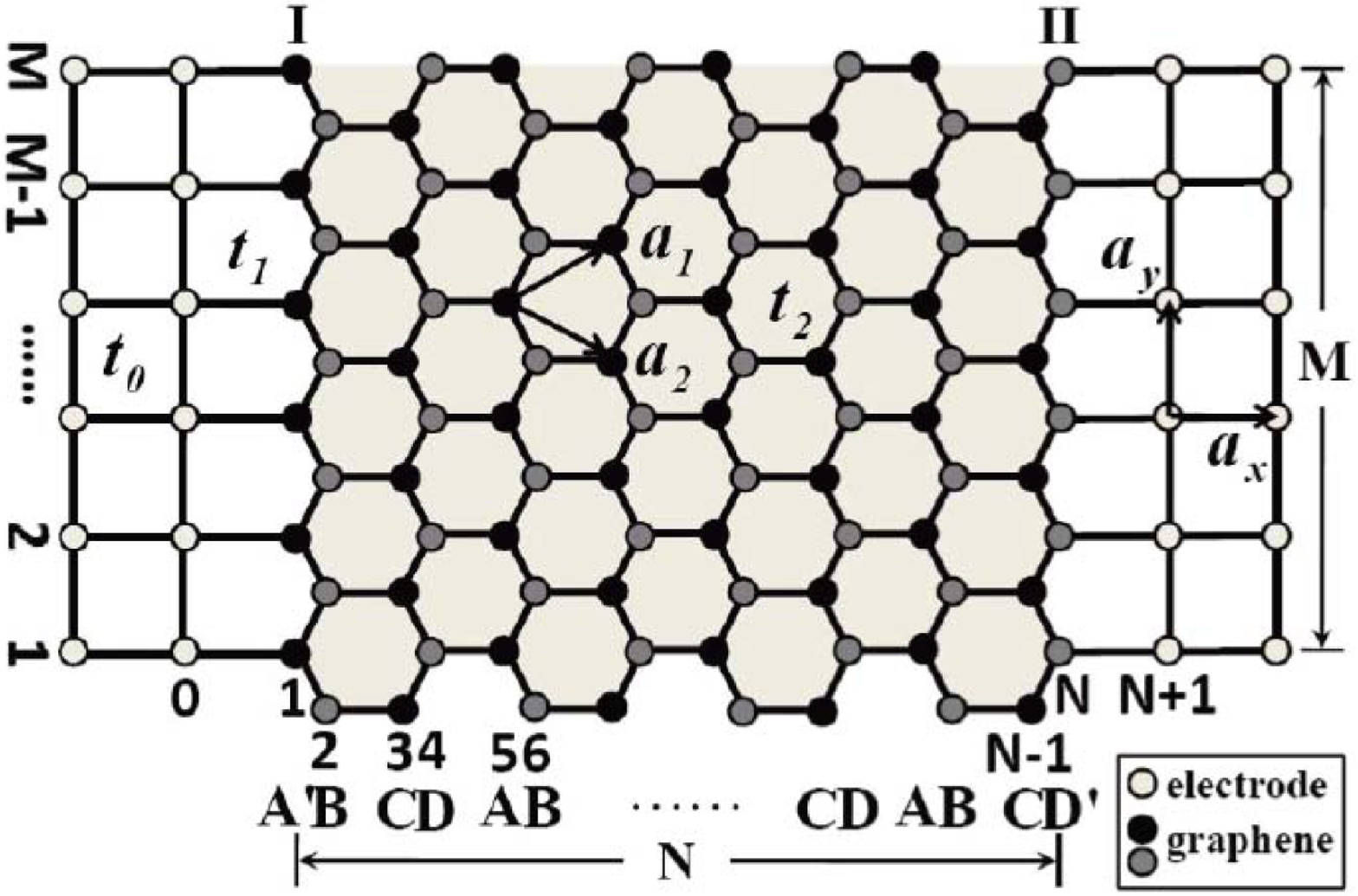}
 \caption{Schematic representation of an armchair graphene
sheet connected to electrodes at interfaces I and II. Primitive vectors
${\bf a}_{x}$ and ${\bf a}_{y}$ for the electrodes and ${\bf a}_{1}$ and
${\bf a}_{2}$ for the sheet give length $L = \sqrt{3} N |{\bf a}_{1}|/4$ and
width $W = (M-1)|{\bf a}_{1}|$; $a_1 = a_2 = a_y$ with $a_1 = 2.46$\AA~for
graphene. The ZGS case, obtained by a $\pi/2$ rotation, has length $L =
(N-1)|{\bf a}_{1}|$ and width $W = \sqrt{3} M |{\bf a}_{1}|/4$, still with
$N \times M$ sites.} \label{nfig1}
\end{figure}

In this paper, we consider graphene sheets of both armchair (AGS) and
zigzag (ZGS) geometry, connected to two metallic leads as illustrated
in Fig.~\ref{nfig1}. By establishing a transfer-matrix formulation within
this minimal model, we present exact results for the anomalous intrinsic
transport properties of graphene. We demonstrate that the AGS and ZGS are
completely different, and explain in detail the non-universal dependence
of $\sigma$ and $F$ on geometry, filling, and gate voltage.

The low-energy properties of graphene can be described by a nearest-neighbor,
one-orbital tight-binding model for $\pi$-electrons on a hexagonal lattice,
\begin{equation}
H = -t_{2} \sum_{\langle ij,i^{\prime}j^{\prime}\rangle} c_{ij}^{\dag}
c_{i^{\prime}j^{\prime}} + \mu \sum_{ij} c_{ij}^{\dag} c_{ij},
\label{eq1}
\end{equation}
where $c_{ij}^{\dag}$ is an electron creation operator at lattice
site ${\bf r}_{ij}\equiv (x_i,y_j)$, $\langle \dots \rangle$ denotes
nearest-neighbor sites, $t_2$ is the hopping integral, and $\mu$ the
chemical potential. The two electrodes are represented by semi-infinite
rectangular strips with hopping $t_0$, while the interface hopping is
$t_1$. We take the interface contact to be perfect and impose open boundary
conditions on the two free edges of the sheet; it is the geometry of these
edges which determines our nomenclature (AGS or ZGS). Because graphene has
two sublattices, sheets of size $N \times M$ lattice sites are taken to have
width $M$ and length $N = 4m$ (AGS) or $N = 2m$ (ZGS) with $m$ an integer.

We begin with the AGS case (Fig.~1) by constructing a
transfer-matrix equation for the scattering of electrons between two
electrodes. In the Schr\"{o}dinger equation $\hat H \psi(E) = E
\psi(E)$, the wave function is represented as $\psi(E) = \sum_{ij}
\alpha_{ij} |ij \rangle$ [$|ij \rangle
 = c_{ij}^{\dag} |0 \rangle$ for ${\bf r}_{ij}$],
 with the complex coefficients
$\alpha_{ij}$ to be determined. $E = E_F$ is the Fermi energy of the
electrodes, which is set by their occupation $n_c$. There are $M$
right- and $M$ left-traveling waves (channels) in each electrode,
each channel characterized by a transverse wavenumber $k_y^n =
\frac{n \pi}{M + 1/2}$ with $n = 1, \dots, M$. The longitudinal
wavenumber $k_{x}^n$ is related to $k_{y}^{n}$ by $E_F = -2 t_0
(\cos k_{x}^n + \cos k_{y}^{n})$.

With a unit-amplitude, right-traveling wave incident on the sheet in
the $n$th channel of the left electrode,
\begin{equation}\label{eq2}
  \begin{array}{ll}
    \alpha^{L}_{ij} = & \sum\limits_{n^{\prime}}
    \left(\delta_{n^{\prime}n}e^{i k_{x}^{n^{\prime}}x_i} + r_{n^{\prime}
     n}e^{-i k_{x}^{n^{\prime}} x_i} \right) \sin(k_{y}^{n^{\prime}} y_j), \\
    \alpha^{R}_{ij}= & \sum\limits_{n^{\prime}} t_{n^{\prime}n} e^{i
    k_{x}^{n^{\prime}} x_i} \sin(k_{y}^{n^{\prime}} y_j),
  \end{array}
\end{equation}
for the left and right electrodes, where $r_{n^{\prime}n}$ and
$t_{n^{\prime}n}$ are respectively reflection and transmission
coefficients from channel $n$ to $n^{\prime}$. For each site ${\bf r}_{ij}$
in the sheet, $\sum_{\tau,\delta} \alpha_{i+\tau,j+\delta} =
{\tilde\mu} \alpha_{ij}$, where $\tau,\delta$ specify the
nearest-neighbor sites of ${\bf r}_{ij}$ and $\tilde \mu = (\mu -
E_F)/t_{2}$. We express the $M$ coefficients $\alpha$ for a given
$i$ as the vector $\vec\alpha_i = \left( \alpha_{i1},...,
\alpha_{iM} \right)^{T}$ in order to connect $\vec\alpha_i$ with its
neighboring slices through the $2M$$\times$$2M$ transfer matrix
${\cal T}_i$,
\begin{eqnarray}\label{eq4}
\left[\begin{array}{l}
  {\vec{\alpha}_{i-1}} \\
  {\vec{\alpha}_{i}}
\end{array}\right]={\cal T}_i \left[\begin{array}{l}
  {\vec{\alpha}_{i}} \\
  {\vec{\alpha}_{i+1}}
\end{array}\right].
\end{eqnarray}
A translation period involves
four different slices (Fig.~1), so
${\cal T}_i$ cycles through the four $2$$\times$$2$ block matrices
\begin{eqnarray}\label{eq5}
A & = & \left[
\begin{array}{cc}
\tilde\mu I & -X^{T} \\
          I &      0 \\
\end{array}
\right],\ \ \ B = \left[
\begin{array}{cc}
\tilde\mu Y\ \ & -Y\ \ \\
             I &  0 \\
\end{array}
\right],\nonumber\\
C & = & \left[
\begin{array}{cc}
\tilde\mu I & -X\ \ \\
          I &  0 \\
\end{array}
\right],\ \ \ D = \left[
\begin{array}{cc}
\tilde\mu Y^{T} & -Y^{T} \\
              I &      0 \\
\end{array}
\right],
\end{eqnarray}
where $X$ is lower-bidiagonal with nonzero elements equal to 1, $Y$
is the inverse of $X$, and $I$ is the identity.

Recursive application of Eq.~(\ref{eq4}) for $N$ slices ($N/4$ translation
periods) relates the coefficients $\alpha^{L}$ and $\alpha^{R}$ at the left
and right interfaces through
\begin{equation}\label{eq6}
\left[
\begin{array}{c}
\vec\alpha^{L}_{0} \\
\vec\alpha^{L}_{1}
\end{array}
\right] =\left[
\begin{array}{cc}
\tilde t_{2} I & 0 \\
             0 & I \\
\end{array}
\right] (ABCD)^{\frac{N}{4}} \left[
\begin{array}{cc}
I & 0 \\
0 & \tilde t_{1} I \\
\end{array}
\right] \left[
\begin{array}{l}
\vec\alpha^{R}_{N} \\
\vec\alpha^{R}_{N+1}
\end{array}
\right], \nonumber
\end{equation}
where $\tilde t_1=t_1^2/(t_0 t_2), \tilde t_2=1/\tilde t_1$.
By considering the one-period transfer matrix $ABCD$,
one finds that the transverse modes in the AGS are unmixed by
scattering processes, remaining independent and retaining the
free-particle dispersion $\epsilon_n = - 2 - 2 \cos k_y^n$. This
makes the LB formalism underlying our transport calculations
particularly appropriate. Thus Eq.~(\ref{eq6}) can be decomposed
into the set of binary linear equations
\begin{equation}\label{eq8}
\!\!\!\! \left[ \!\!
\begin{array}{cc}
- \tilde t_{1} \! & \! g_n \\
\! - e^{-ik^{n}_{x}} \! & \! h_n
\end{array}
\!\! \right] \!\! \left[ \!\!
\begin{array}{c}
r_{nn}\\
t_{nn}
\end{array}
\!\! \right] \!\! = \!\! \left[ \!\!
\begin{array}{c}
\tilde t_{1}\\
e^{ik^{n}_{x}}
\end{array}
\!\! \right] \!\! ,
\left[ \!\!
\begin{array}{c}
g_n \\ h_n
\end{array}
\!\! \right] \!\! = \!\!
\left[ \!\!
\begin{array}{cc}
 a_{n} \! & \! b_{n} \\
-b_{n} \! & \! c_{n} \\
\end{array}
\! \right]  ^{\frac{N}{4}}\! \left[ \!\!
\begin{array}{c}
1\\
\tilde t_{1} e^{ik^{n}_{x}}
\end{array}
\!\! \right] \!\! , \nonumber
\end{equation}
with $a_{n} = (\tilde \mu^{2} - \tilde \mu^{4}) \epsilon_{n}^{-1} -
2 \tilde \mu^{2} - \epsilon_{n}$, $b_{n} = ( \tilde\mu^{3} -
\tilde\mu ) \epsilon_{n}^{-1} + \tilde \mu$, $c_{n} = (\tilde\mu^{2}
- 1) \epsilon_{n}^{-1}$, and $a_{n}c_{n} + b^{2}_{n} = 1$.

Analytic solution for $t_{nn}$ from Eq.~(\ref{eq8}) gives the
transmission probability $T_n\equiv T(k_y^{n})=\left|t_{nn}
\right|^2$ as
\begin{equation}\label{trans}
T_n =
\frac{1}
{\gamma_{1} \cosh \left(N\theta_{n}/2\right) +
\gamma_{2} \sinh \left(N\theta_{n}/2\right) + \gamma_{3}} ,
\end{equation}
where $\gamma_{1} = ( \nu^{2}_{1} + \nu^{2}_{2} + \nu^{2}_{3} +
\nu^{2}_{4} )/8$, $\gamma_{2} ={\rm sign}(\kappa^{2}_{\pm}) (
\nu_{1} \nu_{2} + \nu_{3} \nu_{4} )/4$, $\gamma_{3} = ( -
\nu^{2}_{1} + \nu^{2}_{2} - \nu^{2}_{3} + \nu^{2}_{4} )/8$, $\nu_{1}
= ( -2 \xi_{-} + \xi_{+} \tilde t_{+} \cos k^{n}_{x})/\sin
k^{n}_{x}$, $\nu_{2} = \tilde t_{-} \cos k^{n}_{x} /\sin k^{n}_{x}$,
$\nu_{3}=\xi_{+} \tilde t_{-}$, $\nu_{4}=\tilde t_{+}$, $\xi_{\pm} =
( \kappa_{+}/\kappa_{-} \pm \kappa_{-}/\kappa_{+} )/2$,
$\kappa_{\pm} = ( c_{n} - a_{n} \pm 2b_{n} )^{1/2}$, $\tilde t_{\pm}
= \tilde t_{1} \pm \tilde t_{2}$, and $\cosh \theta_{n} = ( a_{n} +
c_{n} )/2$. The conductivity $\sigma$ and Fano factor $F$ may now
be computed exactly from Eqs.~(\ref{eq8}) and (\ref{trans}), leading to
\begin{equation}\label{land}
\sigma = \frac{\sqrt{3}N}{4M}\frac{2e^{2}}{h} \sum_{n} T_{n},\ \ \
\ F = \frac{\sum_{n} T_{n} ( 1 -  T_{n}) } { \sum_{n} T_{n} }.
\end{equation}
These expressions are completely general within the LB framework, and
are applicable for all sheet sizes $(N,M)$.

For the purposes of this Letter, we focus on the physical insight
contained in Eq.~(\ref{land}) for the situation relevant to most
graphene experiments, namely wide electrodes patterned onto the
sample with rather narrow separation \cite{fano1,fano2}. In this
limit of large $W/L(\sim M/N)$, the sum in Eq.~(\ref{land}) is
replaced by an integral over $k_y$. For convenience we set $t_0 =
t_1 = t_2$, which creates no special symmetries. The Dirac-point
conductivity $\sigma_0$ and the corresponding Fano factor $F_0$ may
then be expressed analytically as
\begin{eqnarray}\label{LB1}
\sigma_0 & = & \frac{e^2}{h} \frac{2\sqrt{3} \text{arctan} \left(\left|
\cos k^{c}_{x}/\sin k^{c}_{x}\right| \right)}{\pi \sin k^{c}_{y} \left|
\cos k^{c}_{x}/\sin k^{c}_{x}\right|},\nonumber \\
F_0 & = & \frac{1}{2} \sec^{2} k^{c}_{x} - \frac{\left| \sin k^{c}_{x}/
\cos k^{c}_{x} \right|}{2\text{arctan}\left(\left|\cos k^{c}_{x}/\sin
k^{c}_{x}\right|\right)},
\end{eqnarray}
where $k^{c}_{y}$ is the Dirac-point wavenumber $2\pi/3$ of the AGS and
$k^{c}_{x}$ is determined by $E_F$. At half-filling of the electrodes,
{\it i.e.}~$n_c = 1$ and $E_F = 0$, we obtain $\sigma_0 = 2e^2/ \sqrt{3}
h \approx 1.1547e^2/h$ and $F_0 = 2 - 3\sqrt{3}/\pi \approx 0.3460$.

Our results for the AGS are similar but not identical to the values
$4e^2/\pi h$ and $1/3$ of Ref.~\cite{Landauer2}. While the electronic
structure of the electrodes leads to a small quantitative difference
between the two studies, we will show below that the symmetry-breaking
effect of the electrode interfaces causes a strong qualitative difference.
The fact that $\sigma_0 \neq 0$ at the Dirac point $\tilde\mu = 0$, despite
the vanishing density of states, is an intrinsic property of the AGS quite
distinct from conventional mesoscopic systems.

\begin{figure}[b]
\includegraphics[width=7.8cm]{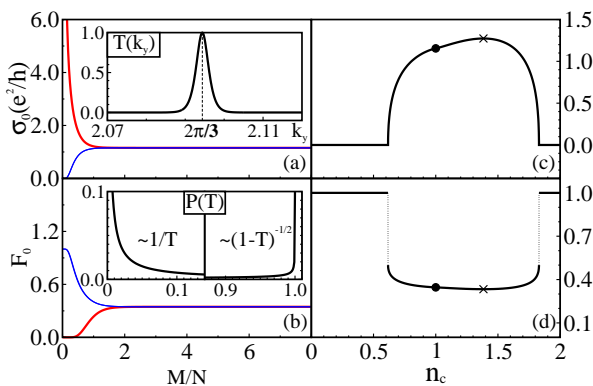}
\vspace{-0.3cm}\caption{(Color online) Dirac-point conductivity $\sigma_0$
(a,c) and Fano factor $F_0$ (b,d) as functions of $M/N$ for $n_c = 1$ (a,b)
and of $n_c$ with $M/N = 10$ (c,d). Dots denote $\sigma_0 = 2e^2/\sqrt{3}h$
in (c) and $F_0 = 2 - 3\sqrt{3}/\pi$ in (d) at $n_c = 1$, while crosses
denote $\sigma_0 = 4e^2/\pi h$ in (c) and $F_0 = 1/3$ in (d), obtained at
$n_c \approx 1.39$. Insets: transmission probability $T(k_y)$ around
$k_y^c = 2\pi/3$ in (a) and its distribution $P(T)$ (see text) in (b).
Calculations performed with $N = 1000$ in the system of Fig.~1.}
\label{nfig2}
\end{figure}

To analyze the physical origin of this behavior, Fig.~\ref{nfig2}
shows the full dependence of $\sigma_0$ and $F_0$ on $n_c$ and on
the aspect ratio $M/N$ of the sheet for $\tilde{\mu} = 0$. In Figs.~2(a)
and (b), for $n_c = 1$, $\sigma_0$ and $F_0$ alternate as $M$ increases
between semiconducting and metallic behavior, the latter obtained
when mod$(2M+1,3) = 0$ and there exists a
resonant channel with $T_n = 1$ at $k_y^n = k_y^c$ [inset Fig.~2(a)].
As the sheet width is increased, the two branches merge at $M/N \sim
1.5$ with $\sigma_0$ and $F_0$ independent of $M/N$; only when $M
\gtrsim N$ are sufficiently many channels with $T_n \sim 1$
available that their contributions to the sum in Eq.~(\ref{land})
are constant. For such sheets [inset Fig.~2(b)], channels with $T
\gtrsim 0.9$ contribute to $F_0$ with a distribution $P(T) \propto
1/\sqrt{1 - T}$ while channels with $T \lesssim 0.1$ have $P(T)
\propto 1/T$ \cite{prob}. Although $P(T)$ resembles the universal
bimodal distribution for a disordered mesoscopic system, which also
has $F_0 \sim 1/3$, the underlying physics is completely different:
the sub-Poissonian behavior is caused by the interference of
relativistic quantum particles, which results in transport
contributions $T_n \sim \exp(- |k_y^n - k_y^c| N)$ away from the
resonant channel [inset Fig.~2(a)]. This type of behavior, namely
$\sigma_0 \neq 0$ and $F_0 \sim 1/3$, is obtained in the AGS only
for $0.63 \lesssim n_c \lesssim 1.83$ [Figs.~2(c) and (d)].

Figure 3 shows the effects of a gate voltage on $\sigma$ and $F$ for
$n_c = 1$. For a finite sheet, the number of Fermi momenta (number
of energy bands intersected) increases with $\mu$, each peak in
$\sigma$ and $F$ corresponding to one more contributing resonant
channel. When the sheet is sufficiently wide [Fig.~3(a)], channels
are added at nearly equal intervals, resulting in almost periodic
oscillations, whereas for $M/N \lesssim 1.5$ [Fig.~3(b)] the effects
of added channels appear quasi-periodic. Superposed on the
oscillation is a linear and slightly asymmetric behavior of $\sigma$
about the Dirac point. The former is a consequence of the linear
dispersion of graphene and the latter of the electron-hole asymmetry
caused by the electrodes \cite{Beenakker}. Thus our exact results
illustrate the inherent dependence of experimental observations on
both $W/L$ and $L$ \cite{exp1,exp3}, and demonstrate further that
such behavior can be intrinsic, rather than appearing only as a
consequence of sample disorder or interfacial defects.

\begin{figure}[b]
\includegraphics[width=8cm]{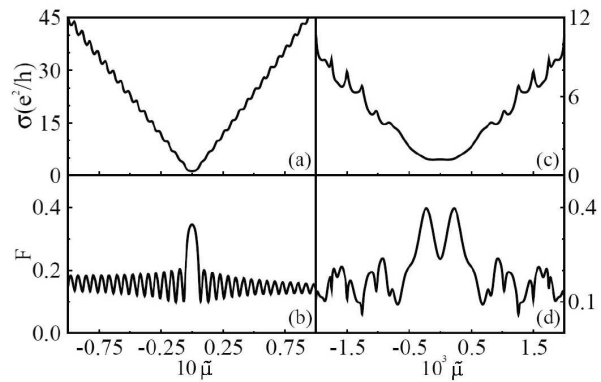}
\caption{Conductivity $\sigma$ (a,c) and Fano factor $F$ (b,d) for the AGS
with $M = 10000$ and $n_c = 1$, shown as functions of ${\tilde \mu}$ for
$M/N = 10$ (a,b) and $M/N = 1$ (c,d).} \label{nfig3}
\end{figure}

We turn now to the ZGS. The geometry of this case requires a
transfer matrix ${\cal T}_i$ expressed in terms of two 2$\times$2
block matrices and a quartic form of Eq.~(\ref{eq8}) for $t_{nn}$,
which is solved numerically to obtain $\sigma$ and $F$ from
Eq.~(\ref{land}). Figure 4 shows the dependence of $\sigma_0$ and
$F_0$ on $M/N$, again with $t_0=t_1 = t_2$ and $\tilde{\mu}= 0$. The
ZGS also possesses metallic and semiconducting branches, which
alternate with respect to the sheet length $N$ rather than to its
width $M$. The asymptotic behavior is metallic, with $\sigma_0^u
\propto 16/\sqrt{3} \frac M N$ for mod$(2N+1,3) = 0$, and
semiconducting with $\sigma_0^d \sim 0.2801/N^2$ otherwise
[Fig.~4(a)]. The corresponding Fano factors [Fig.~4(b)] are $F_0^u
\sim (1 - 0.2801/N^2)/(1+ 32.98N^2/\frac M N)$ and $F_0^d \sim 1 -
0.08223/N^2$, the two branches merging only when $\frac M N\gg
32.98N^2$. Graphene sheets in this limit of $W/L$ would therefore
have $\sigma_0 = 0$ and $F_0 = 1$ at the Dirac point, implying a
Poissonian shot-noise quite different from the AGS. A finite minimal
conductivity, $\sigma_0 \lesssim 2 e^2 / \sqrt{3} h$ (reaching its
maximal value when $E_F = \pm 1$), and a sub-Poissonian $F_0$ are
obtained for all $n_c \neq 0, 1, 2$ [Figs.~4(c) and (d)].

The origin of the contrasting intrinsic transport properties of the
AGS and ZGS for the Dirac point lies in the special nature of zigzag
chains in graphene. The key point is how this affects scattering at
the interfaces. Because $\sum_{\tau,\delta} \alpha_{i+\tau,j+\delta}
 = 0$ for any site ${\bf r}_{ij}$ in the sheet, the wavenumber of extended
states is $2\pi/3$ when projected onto the zigzag chain direction, and
zero in the orthogonal direction. In the AGS, zigzag chains are parallel
to the interfaces so that $k_y^c = 2\pi/3$ for mod$(2M+1,3) = 0$. Thus
the incident traveling wave is not deformed at the interface and there
is no interfacial scattering. Consequently, $T(k_y^c) = 1$ and $T(k_y^n)
\sim 1$ in a regime of width $O(1/L)$ about $k_y^c$ [inset Fig.~2(a)],
resulting in a finite $\sigma_0$ after the summation in Eq.~(\ref{land})
if $0.63 \lesssim n_c \lesssim 1.83$ [Fig.~2(c)]. In the ZGS, zigzag chains
connect the left and right electrodes, $k_y^c = 0$, and the armchair
interfaces involve two sublattices, with two values of $k_x^n$
corresponding to each $k_y^n$. This induces interfacial scattering.
As a consequence, for $n_c = 1$ the transmission amplitudes are
suppressed very strongly for any $k_y^n \neq k_y^c$ and $T_n = \delta_{k^n_y,
k^c_y} + T_{b}$ [inset Fig.~4(a)], where $T_b$ is a very small background
of width $O(1/L^3)$ arising from interfacial scattering. Neither term
contributes to the integral in the limit of large $W/L$ and $L$, whence
$\sigma_0 = 0$ and $F_0 = 1$. When $n_c \neq 1$ \cite{details}, imaginary
$k_x^n$ values appear for some channels $\{k_y^n\}$, giving contributions
to $T(k_y^n)$ over a greater width and leading to a finite $\sigma_0$
[Fig.~4(c)]. Thus it is the topological difference in the geometry
along and across a hexagonal lattice which results in two fundamentally
different types of interfacial scattering, and hence in the contrasting
intrinsic transport properties of AGS and ZGS systems. This microscopic
insight was not included in any previous studies.

\begin{figure}[t]
\includegraphics[width=7.8cm]{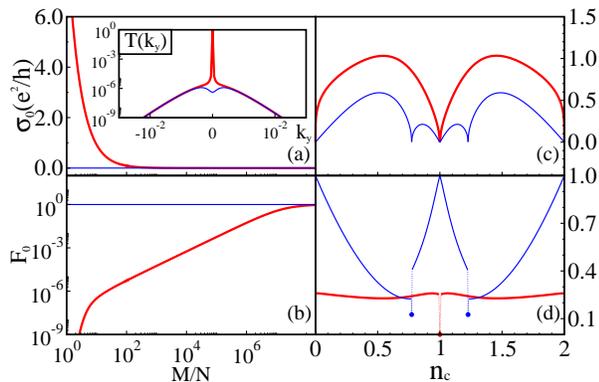}
\caption{(Color online) $\sigma_0$ (a,c) and $F_0$ (b,d) for a ZGS, shown as
functions of $M/N$ at $n_c = 1$ (a,b) and of $n_c$ for $M/N = 1200$ (c,d).
Inset: $T(k_y)$ around $k_y^c = 0$. Red and blue curves indicate respectively
metallic and semiconducting situations, calculated with $N = 1000$ and 999.}
\label{nfig4}
\end{figure}

Many investigations of graphene transport may be found in recent
literature. Augmenting the general results cited above, experimental
studies of the conductivity minimum have addressed the coherence of
Dirac-point transport \cite{rhea}, the role of contacts and sample
edges \cite{rlea}, and how interface charging leads to asymmetric
gate-voltage effects \cite{rhea2}. Many theoretical studies have
considered transmission coefficients in a finite graphene system,
all restricted (as here) to the case of non-interacting electrons:
from its weak interactions and the vanishing density of states at
the Dirac point, the fundamental transport properties of graphene
are expected to emerge at the band-structure level. These
investigations all differ from ours in the approximations applied,
or in system size and geometry, or in the method of calculation, and
hence in the nature of their conclusions. In an effective contact
model \cite{rs} for a sufficiently large system, mode selection at
the Dirac point makes all leads equivalent. Numerical treatments, of
the same model \cite{rrs} and in a more general framework
\cite{rblvkc}, have probed size, gate-voltage, and impurity effects.
While these and other studies \cite{Landauer2,rbm} note that the AGS
and ZGS cases should differ, the fundamentally different nature
($\sigma_0 = 0$, $F_0 = 1$) of the ZGS case and the microscopic
origin of the different intrinsic transport properties have been
missed. Further, because we have analyzed the intrinsic transport
arising due to lead and interface geometry, we may conclude that
disorder effects are not required to obtain the anomalous behavior
observed in experiment \cite{exp1,exp3,rhea}.

To conclude, we have presented exact solutions of the
transfer-matrix equations for graphene sheets with metallic
electrodes. Our results are microscopic and completely general, and
can be used to show that the Dirac-point conductivity and the Fano
factor tend respectively to $\sigma_0
 = 2e^2/\sqrt{3} h$ and $F_0 = 2 - 3\sqrt{3}/\pi$ for armchair graphene
sheets in the short and wide limit relevant to experiments. The same
quantities tend to 0 and 1 respectively for zigzag graphene sheets.
Our exact results suggest that the measured finite minimum
conductivity and sub-Poissonian Fano factor are the consequence of
armchair rather than zigzag graphene systems, and show how this
fundamental difference depends on the availability of resonant
transmission channels, which is determined in turn by the geometry
of the hexagonal lattice.


The authors thank B. Normand, E. Tosatti, B. G. Wang, X. R. Wang, X. C.
Xie, Lu Yu, and Y. S. Zheng for fruitful discussions. This work was
supported by the Chinese Natural Science Foundation, Ministry of Education,
and National Program for Basic Research (MST).


\begin{thebibliography}{99}


\bibitem{exp0} K. S. Novoselov {\it et al.}, Science {\bf 306}, 666 (2004).

\bibitem{fermion} E. Fradkin, Phys. Rev. B {\bf 33}, 3263 (1986);
N. H. Shon and T. Ando, J. Phys. Soc. Japan {\bf 67}, 2421 (1998);
E. V. Gorbar, V. P. Gusynin, V. A. Miransky, and I. A. Shovkovy,
Phys. Rev. B {\bf 66}, 045108 (2002).



\bibitem{exp1} K. S. Novoselov {\it et al.}, Nature {\bf 438}, 197 (2005).

\bibitem{exp3} F. Miao {\it et al.}, Science {\bf 317}, 1530 (2007).


\bibitem{Landauer1} M. I. Katsnelon, Euro. Phys. J. B {\bf 51}, 157 (2006).

\bibitem{Landauer2} J. Tworzydlo {\it et al.}, Phys. Rev. Lett. {\bf 96},
246802 (2006).



\bibitem{fano1} R. Danneau {\it et al.}, Phys. Rev. Lett. {\bf 100},
196802 (2008).

\bibitem{fano2} L. DiCarlo, {\it et al.}, Phys. Rev. Lett. {\bf 100},
056801 (2008).


\bibitem{prob} The distribution of the transmission probability is defined as
$P(T) = \frac 1\pi \partial k_y/\partial T$.

\bibitem{Beenakker} When graphite leads are used, $\sigma$ is symmetric
\cite{Landauer2}.



\bibitem{rhea} H. B. Heersche {\it et al.}, Nature {\bf 446}, 56 (2007).

\bibitem{rlea} E. J. H. Lee {\it et al.}, Nature Nanotech. {\bf 3},
486 (2008).

\bibitem{rhea2} B. Huard, N. Stander, J. A. Sulpizio, and D. Goldhaber-Gordon,
Phys. Rev. B {\bf 78}, 121402 (2008).



\bibitem{rs} H. Schomerus, Phys. Rev. B {\bf 76}, 045433 (2007).

\bibitem{rrs} J. P. Robinson and H. Schomerus, Phys. Rev. B {\bf 76}, 115430
(2007).

\bibitem{rblvkc} S. Barraza-Lopez, M. Vanevi\'c, M. Kindermann, and M. Y.
Chou, Phys. Rev. Lett. {\bf 104}, 076807 (2010).

\bibitem{rbm} Y. M. Blanter and I. Martin, Phys. Rev. B {\bf 76}, 155433
(2007).

\bibitem{details}  Detailed analysis for $n_c \ne 1$ will be presented
elsewhere.

\end{thebibliography}
\end{document}